\documentclass[conference]{IEEEtran}
\IEEEoverridecommandlockouts
\usepackage{cite}
\usepackage{amsmath,amssymb,amsfonts}
\usepackage{algorithmic}
\usepackage{graphicx}
\usepackage{textcomp}
\usepackage{xcolor}
\usepackage{url}
\usepackage{booktabs}
\usepackage{tabularx}
\usepackage{datetime2}
\usepackage{hyperref} \hypersetup{hidelinks}
\usepackage{multirow}
\usepackage{array}
\usepackage{orcidlink}
\newcolumntype{L}[1]{>{\raggedright\arraybackslash}p{#1}}
\newcolumntype{Y}{>{\raggedright\arraybackslash}X}

\begin{document}

\title{Janus Vehicles: Entity-Centric Threat Modelling of Protocol-Conformant Vehicles under a Malicious-Manufacturer Assumption}

\author{
	\IEEEauthorblockN{Jaeung Franciskus Yoo\,\orcidlink{0009-0004-1136-1693}}
	\IEEEauthorblockA
	{
		\textit{Korea University} \\
		Seoul, Republic of Korea \\
		franciskus@korea.edu
	}
	\and
	\IEEEauthorblockN{Huy Kang Kim\,\orcidlink{0000-0002-0760-8807}}
	\IEEEauthorblockA
	{
		\textit{Korea University} \\
		Seoul, Republic of Korea \\
		cenda@korea.ac.kr
	}
	\thanks{This is a preprint of a paper accepted for publication in the proceedings of the \textit{31st IEEE Pacific Rim International Symposium on Dependable Computing (IEEE PRDC 2026)}.}
}

\maketitle

\begin{abstract}
Automotive cybersecurity commonly assumes benign end-product manufacturers and design authorities, focusing on supply-chain threats, external attacks, and compromised network participants observable through misbehaviour.
This assumption overlooks malicious-by-design vehicles that remain authentic and protocol-conformant while performing undeclared sensing, communication, or context-triggered actuation.
We formulate an \textit{Entity-Centric Threat Model} spanning three subcomponents.
The first, \textit{Malign Entity Archetypes}, describes four types of malign entity: peer-benign, purpose-malicious vehicles; compromised vehicles; reappropriated genuine components; and imitated vehicles or components.
The model exposes two assurance gaps: 
authenticity does not establish declared-purpose conformance, while malicious functionality can evade detection based on behavioural history or peer-visible evidence.
Reappropriated components can remain genuine and retain usable identities or capabilities outside authorised contexts, reducing the effectiveness of signal fingerprinting and certificate-based authentication.
To structure this threat space, we introduce a \textit{Behaviour--Locus Matrix} that maps covert, overt, and concealment actions across hardware, software, and network loci.
For peer-benign, purpose-malicious vehicles, we further present a \textit{Behavioural Interaction Model} that represents how hidden capabilities are contextually activated, realised through actions, manifested within the vehicle and its environment, and potentially observed or concealed.
\end{abstract}


\section{Introduction}
\label{sec:introduction}
Modern cyber-physical systems (CPS), such as vehicles, are networked multimodal systems that integrate sensing, computation, decision making, communication, and physical actuation.
Accordingly, automotive cybersecurity research has examined attacks against in-vehicle networks, vehicle software, external interfaces, and vehicle-to-everything (V2X) communication.
These studies generally assume that the delivered vehicle was designed to perform its declared function, although it may subsequently be compromised.
Security mechanisms are therefore designed primarily to protect vehicles from external adversaries or to detect compromised vehicles attacking other participants through observable network interactions \cite{AutomotiveSecuritySurvey1, VANETSecuritySurvey1, VANETSecuritySurvey2Trust}.

Because of this benign-design assumption, current automotive cybersecurity pays less attention to hidden functionality deliberately incorporated by a manufacturer or design authority, which may also retain the ability to deploy such functionality later through maintenance operations.
The Volkswagen emissions scandal provides a concrete precedent for context-sensitive deception by an established manufacturer:
defeat-device software detected regulatory testing and altered vehicle behaviour accordingly \cite{UKParliamentVW2016, VWFiatBoschCase}.
Furthermore, related investigations also uncovered emissions-related misconduct by Fiat Chrysler Automobiles (FCA), whose vehicles used the same Bosch-supplied module \cite{VWFiatBoschCase}.
Although the case concerned emissions compliance, it demonstrates that specific software included by design authority can condition its visible behaviour on the assurance environment.
Similarly, a malicious-by-design vehicle would authenticate correctly, transmit valid V2X messages, and remain indistinguishable from benign vehicles to conventional network monitoring while collecting, retaining, or exfiltrating information about its encompassing environments.
The purpose violation need not manifest as an attack on a peer node.
It can instead occur through physical sensing or actuation, an undisclosed communication channel, or misuse of an otherwise legitimate channel such as connected car functionality.

This concern is amplified by the scale of contemporary vehicle data collection.
The BBC reports that modern vehicles collectively generate data across trillions of miles of driving, illustrating the scale at which information about vehicles, occupants, and their surroundings may be collected \cite{bbc_car_spying_2026}.
Automotive privacy concerns have also prompted investigations into Hyundai and Toyota in Australia \cite{ChoiceToyotaHyundaiPrivacy2026}.
Related security concerns have led to US restrictions involving Chinese robots and power inverters \cite{BBCChineseRobotBan2026}, restrictions on Chinese vehicles at UK military sites \cite{TelegraphChineseCarsUKSF2026}, and investigations into Chinese-made buses in Norway, Denmark, and the UK \cite{RuterElectricBusSecurity2025,GuardianChineseBuses2025,YTFChineseBuses2026}.
In a more extreme case, explosive-laden pagers entered circulation in Lebanon, possibly through a shell company operating under a brand licensing arrangement, and were remotely detonated upon receipt of a specific message \cite{BBC2024LebanonPager1, BBC2024LebanonPager2}.
Unlike this deliberate operation, the other cases cited above do not uniformly establish malicious conduct, as several concern ongoing investigations or precautionary restrictions.
Collectively, however, they demonstrate that vendor-controlled data access and remote functionality are treated as credible security concerns across automotive and adjacent CPS domains.

We refer to this threat class as a \textbf{\textit{peer-benign, purpose-malicious}} entity:
an entity that conforms to the protocols governing its legitimate interactions while containing functionality that violates its declared or authorised purpose.
This distinction separates \textbf{protocol conformance} from \textbf{purpose conformance}:
\begin{itemize}
	\item A \textbf{protocol-conformant} vehicle produces interactions that appear syntactically and semantically valid and cause no observable disturbance to nearby entities or networks.
	\item A \textbf{purpose-conformant} vehicle additionally restricts its sensing, storage, communication, and actuation to behaviour authorised for its stated operational role. It truthfully represents its operation, performance, and compliance, and contains no hidden functionality that causes harm or initiates undeclared actions when triggered.
\end{itemize}
\textit{Ipso facto}, protocol conformance is necessary for interoperability, but is not sufficient evidence that the entity as a whole is trustworthy.
Despite the problem, investigation of cyber-physical systems made with clandestine objectives presents practical predicaments.
As described in \cite{VWFiatBoschCase} and \cite{UKParliamentVW2016}, the complexity of vehicles' internal systems and software poses substantial challenges for investigation. 
Black-box testing is particularly difficult because investigators lack access to the internal architecture and implementation details.
Requiring full disclosure of software and hardware designs could raise intellectual property concerns without guaranteeing that the disclosed designs accurately reflect the deployed product.
A sanitised specimen may be submitted for homologation, with furtive functionality introduced later through software updates or maintenance.
Even when such functionality is present, dynamic testing may fail to reveal it if activation depends on a rare combination of contextual conditions.
A system capable of recognising investigative conditions may also destroy evidence, disable itself, or overwrite its malicious code with a benign substitute.
These challenges call for an investigative approach that combines available design and software evidence with runtime, contextual, communication, and physical observations.

This paper is motivated by the following research question:
\begin{quote} \textit{How can investigators characterise and seek evidence of hidden, purpose-violating functionality in a cyber-physical entity that remains conformant at its legitimate protocol interfaces?} \end{quote}
A complete answer requires both domain knowledge about the threat and operational knowledge about how relevant evidence should be acquired and evaluated.
Ontology engineering similarly distinguishes the explicit representation of domain concepts and assumptions from the operational knowledge used to perform a task \cite{Ontology101}.
Following this methodological distinction, this paper addresses the conceptual prerequisite for investigation rather than proposing an investigation model or procedure.
Consequently, evidence acquisition procedures and empirical validation of an investigation method remain outside the scope of this paper.

This paper makes the following contributions:
\begin{itemize}
	\item We formulate \textbf{Malign Entity Archetypes} encompassing malicious-by-design vehicles, benign vehicles under conventional attack, reappropriated genuine components used for misdeeds, and imitated vehicles or components.
	\item We introduce a \textbf{Behaviour--Locus Matrix} that cross-classifies covert, overt, and concealment actions across hardware, software, and network loci. 
	The matrix provides a systematic means of mapping how malign functionality may manifest within a vehicle and what evidence, if any, may be observable to peers or investigators.
	\item For peer-benign, purpose-malicious vehicles, we present a \textbf{Behavioural Interaction Model} that represents vehicle capabilities, contextual triggers, and their realisation through actions.
	It complements the \textit{Behaviour--Locus Matrix} by relating capabilities and activation conditions to realised actions, operational loci, artefacts, investigation, concealment, and declared purpose.
\end{itemize}
Taken together, these three components constitute the proposed \textit{Entity-Centric Threat Model}, addressing gaps in current automotive-security research and regulation.

% %------% %------% %------% %------% %------% %------% %------% %------% %------% %------% %------% %------% %------% %------% %------% %------% %------% %------% 
% %------% %------% %------% %------% %------% %------% %------% SECTION ZWEI HINTERGROUND BEGINS  %------% %------% %------% %------% %------% %------% 
% %------% %------% %------% %------% %------% %------% %------% %------% %------% %------% %------% %------% %------% %------% %------% %------% %------% %------% 
\section{Background}

\begin{figure*}[t]
	\centering
	\includegraphics[width=\textwidth]{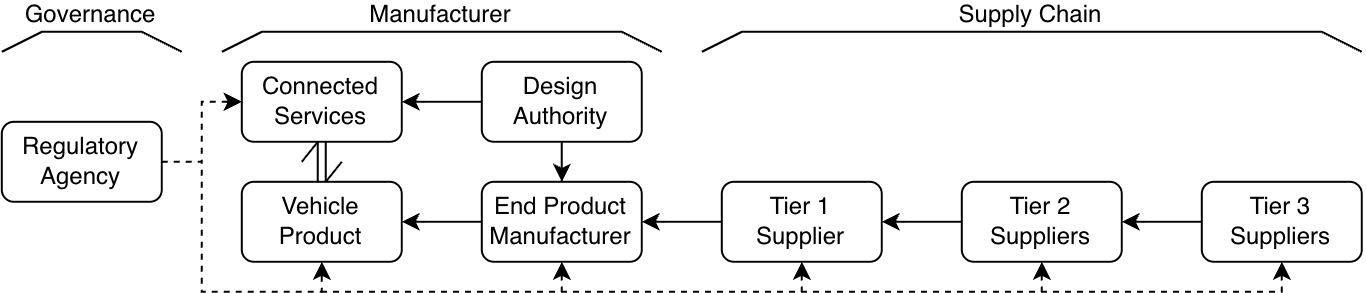}
	\caption{Trust and assurance relationships in the vehicle product ecosystem. Existing regulation addresses governance on connected car applications and supply chain risk but does not explicitly consider a malicious design authority within the manufacturer domain.}
	\label{fig:trust-chain}
\end{figure*}

\subsection{Current Practice in Automotive Security}
Existing vehicular security research mainly concerns protecting vehicles and vehicular networks from compromised or imitated participants, components and vehicle nodes.

\paragraph{Automotive Risk Management}
Automotive threat analysis and risk assessment methods systematically identify assets, damage scenarios, threat scenarios, attack paths, potential impacts, and attack feasibility \cite{ISO21434LandingPage, Heavens2RiskAssessModel}. 
Their primary purpose, however, is prospective cybersecurity risk assessment throughout the vehicle lifecycle, rather than the behavioural investigation of a deployed entity suspected of containing undeclared functionality.
Put differently, these methods principally treat the vehicle as a system to \textbf{\textit{be protected}} from adversaries, not a system to protect the public \textbf{\textit{from}}. 

\paragraph{Supply Chain Security in Automotive Industry}
Supply chain cybersecurity guidance partially relaxes this benign design assumption.
\textit{NIST SP 800-161 Rev. 1} explicitly recognises that acquired products and services may contain malicious functionality and provides enterprise level practices for identifying, assessing, and mitigating such risks \cite{NIST-C-SCRM}.
\textit{MITRE}'s \textit{System of Trust} similarly evaluates risks associated with suppliers, supplies, and services \cite{MITRESoT}.
\textit{Auto-ISAC}'s \textit{TRACK} Programme applies this supplier assurance perspective to the automotive ecosystem by supporting common supplier assessments and OEM visibility into third party cybersecurity risk \cite{AutoISACTRACK2026}.
These frameworks generally position a well-intentioned acquiring organisation, including a manufacturer, as the party assessing and managing risks arising from suppliers, products, services, and other dependencies.
Yet, they do not directly address the inverse trust relationship, in which the manufacturer or design authority itself may pose a risk to users, regulators, and other system participants.
This leaves a downstream assurance question:
whether users, regulators, and other system participants can trust the operational purpose of a finished entity when its design authority may itself be adversarial.

\paragraph{Behaviour-Based Automotive Security}
Conventional automotive cybersecurity research, particularly work on vehicular networks, commonly addresses vehicle misbehaviour and the trustworthiness of participating nodes.
Vehicular misbehaviour detection considers faulty or malicious participants in a network, including authenticated insiders, but primarily seeks evidence of observable interaction and information exchanged through cooperative vehicular networks.
Such detectors seek indicators including implausible messages, identity inconsistencies, protocol violations, or deviations from expected behaviour.
Existing approaches commonly evaluate message plausibility, sender behaviour, reputation, or consistency with information received from other participants\cite{VANETSecuritySurvey1, VANETSecuritySurvey2Trust, MisbehaviourDetectionSurvey}.

Consequently, a vehicle that remains conformant at its cooperative interfaces while performing clandestine actions such as promiscuous sensing, out-of-band exfiltration, or dormant local actuation may produce none of the evidence assumed by the detectors.
Furthermore, behavioural detection and trust-management systems are inherently \textit{reactive} and may therefore fail to prevent a one-time attack.
Although VANET taxonomies also recognise passive attackers, passive behaviour in this context generally denotes observation or eavesdropping on a network~\cite{VANETSecuritySurvey1}.
It does not capture a protocol-conformant vehicle that has malicious or undeclared objectives.
Such a vehicle therefore may produce little or none of the evidence assumed by communication or behaviour-oriented detectors.

\paragraph{Prevalent Regulation in Automotive Industry}
Automotive cybersecurity governance combines homologation instruments with broader organisational and product-level obligations.
\textit{ISO/SAE 21434} defines lifecycle cybersecurity engineering processes for road vehicle electrical and electronic systems \cite{ISO21434LandingPage}.
\textit{UN Regulation No. 155} establishes \textit{Cyber Security Management System} and vehicle type cybersecurity requirements, while \textit{UN Regulation No. 156} establishes \textit{Software Update Management System} and software update requirements \cite{UNR155, UNR156}.
Beyond homologation, the \textit{NIS2} Directive imposes cybersecurity risk management and incident reporting obligations on qualifying motor vehicle manufacturers under \textit{NACE Rev. 2 Section C, Division 29} \cite{NIS2Directive}.
The \textit{Cyber Resilience Act} (CRA) imposes lifecycle cybersecurity and vulnerability handling requirements on products with digital elements \cite{EUCRA2024}.
Although vehicles covered by sector specific type approval are excluded, the CRA can still apply to automotive digital products marketed separately. \\
\indent Taken together, these instruments broaden cybersecurity oversight but generally rely on manufacturers or economic operators to identify and mitigate threats.
Figure~\ref{fig:trust-chain} shows the resulting gap: 
current governance and research do not explicitly establish what must be assured when the design authority itself is potentially adversarial.
Fraudulent conduct or breaches of adjacent law may still incur consequences, but homologation regulation lacks an explicit standard and process for providing such assurance.
Importantly, the design authority may differ from the manufacturer or brand owner, leaving those parties dependent on its technical representations.
The 2016 Mitsubishi fuel-consumption testing scandal illustrates this separation, as Mitsubishi-designed vehicles were sold under both Mitsubishi and Nissan brands \cite{MitsubishiFuelTesting2016}.
Unlike the Volkswagen scandal, scrutiny in this case was initiated by Nissan after it identified an anomaly in the product.
Current instruments require risk management, documentation, testing, and incident response, but do not explicitly require assurance that all implemented functionality is declared and serves an authorised purpose.
They therefore do not directly address a design authority that conceals dormant or trigger-activated, purpose-violating functionality within a genuine vehicle.

\subsection{Behaviour-Centric Malware Characterisation}
\label{sub:malwareBehaviour}
Malware provides a useful conceptual precedent because it is intentionally engineered to pursue an adversarial objective and may conceal its capabilities behind apparently legitimate operation.
Malware analysis therefore treats the examined artefact as potentially adversarial and seeks evidence of concealed capabilities and behaviours.
This perspective can inform the investigation of malicious-by-design cyber-physical entities.

Malware detection commonly encompasses signature, heuristic, and behaviour based approaches~\cite{malwareAPI, malwareSurveyAccess}.
Signature-based methods identify patterns or indicators derived from previously characterised specimens.
For malicious-by-design vehicles, however, an investigator cannot generally assume access to a comparable collection of representative samples.
Vehicle behaviour may also be distributed across vendor-specific hardware, software, communication systems, and physical components, and may emerge only under particular contextual conditions.
Exact pattern matching alone is therefore unlikely to provide sufficient investigative coverage, whereas behaviour-centric analysis offers a more transferable analytical perspective.

Behavioural frameworks provide vocabularies and representations for recording the resulting findings.
The \textit{Malware Behavior Catalog} (MBC) organises malware behaviours according to objectives such as collection, command and control, exfiltration, impact, and defence evasion~\cite{MitreMBC}.
\textit{Malware Attribute Enumeration and Characterization} (MAEC) provides a structured means of representing malware capabilities, behaviours, actions, affected objects, observables, and their relationships~\cite{MitreMAEC5}.
These approaches shift the investigative question from whether an artefact matches a known signature to what it is capable of doing and what evidence supports that assessment.

\begin{figure*}[t]
	\centering
	\includegraphics[width=\textwidth]{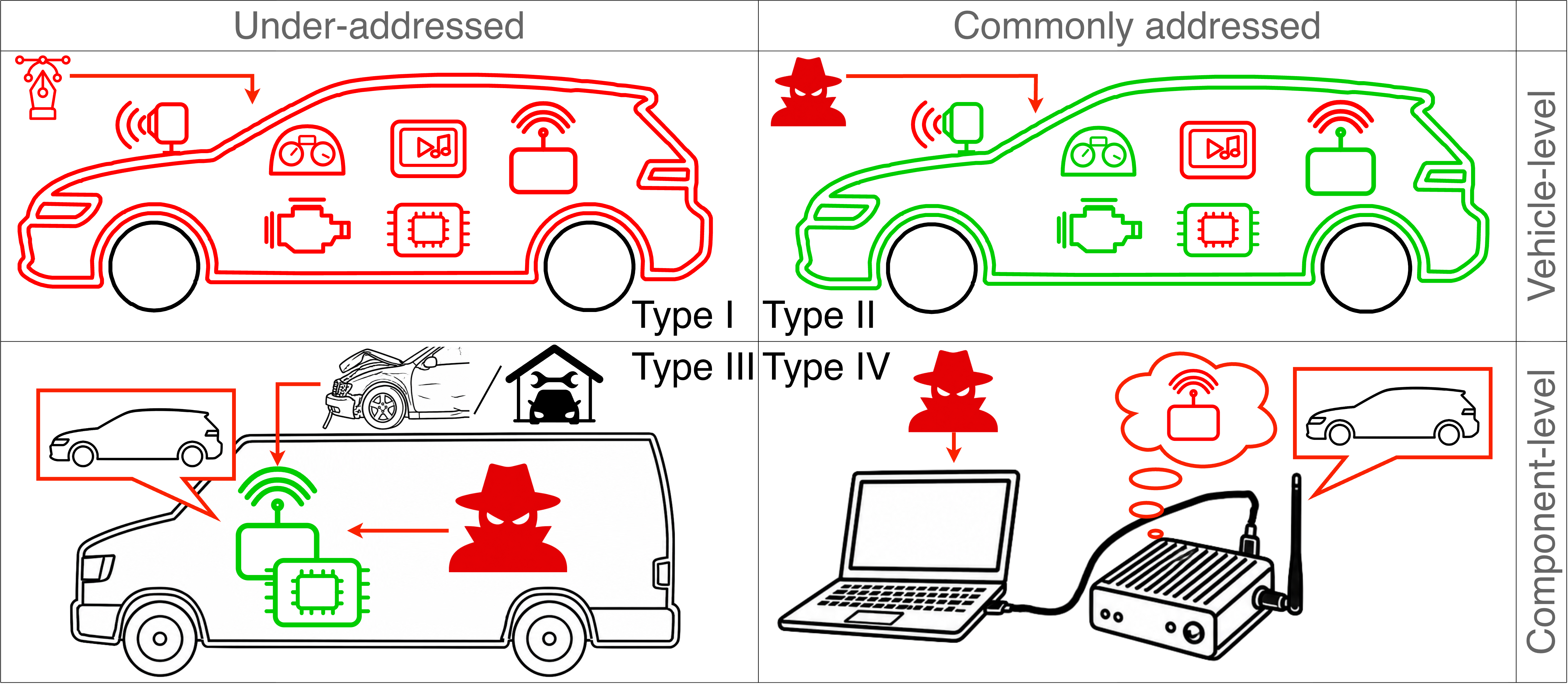}
	\caption{	Four \textit{Malign Entity Archetypes}, organised by entity level and the extent to which addressed in existing automotive security research:
			(I) malicious-by-design vehicle,
			(II) benign vehicle under conventional attack,
			(III) reappropriated genuine component, and
			(IV) imitated vehicle or component.}
	\label{fig:fourThreatTypes}
\end{figure*}

Although behaviour-centric malware analysis provides a useful conceptual foundation, it cannot be transferred unchanged to vehicles.
Unlike a reproducible software specimen, a vehicle is a distributed CPS whose behaviour depends on interactions among software, hardware, networks, physical components, its environment, and potentially external services.
Restoring an individual software component may therefore neither reproduce the vehicle's distributed state nor prevent investigation-aware concealment through altered behaviour, erased records, or manufacturer notification.
Its adaptation consequently requires contextual stimulation, observation across cyber-physical loci, and evidence sources independent of potentially adversarial internal mechanisms, motivating the threat model developed in the following section.

%--- SECTION 3 ---
\section{Entity-Centric Threat Model}
To establish a conceptual foundation for the threat enabled by this gap, we propose an \textbf{Entity-Centric Threat Model} for vehicular security. 
The model comprises three components: 
\textit{\nameref{sec:fourthingies}}, 
\textit{\nameref{sec:blm}}, and 
\textit{\nameref{sec:bim}}.

\subsection{Malign Entity Archetypes}
\label{sec:fourthingies}
We distinguish four illustrative adversary configurations according to the origin of malicious functionality, the authenticity of the resulting entity, and the extent of adversarial control.
Figure~\ref{fig:fourThreatTypes} provides an overview of these configurations.

\paragraph{Type I: Malicious-by-design vehicle}
A manufacturer or equivalent design authority intentionally incorporates hidden malicious functionality into a genuine vehicle product.
The vehicle is neither impersonated nor modified by an external attacker: its hardware, firmware, software, credentials, and identity are authentic according to the manufacturer's own trust infrastructure.
Because the adversary participates in system design, malicious functionality can be coordinated across multiple components and operational layers.

Such functionality is particularly difficult to detect when it remains passive and covert.
The Volkswagen emissions scandal illustrates the corresponding investigative difficulty: software implemented by the manufacturer distinguished laboratory testing from normal driving and altered vehicle behaviour accordingly, yet the UK Transport Committee concluded that existing approval authorities and technical services would not have revealed the deception \cite{UKParliamentVW2016}.

Furthermore, the manufacturer can design sophisticated monitoring and concealment mechanisms capable of recognising investigative conditions and hiding the relevant functionality.
This design level control enables both coherent coordination among dedicated malicious functions and additional protective layers that conceal them, increasing the difficulty of detection.
For example, the existence or purpose of certain messages may be concealed through non-standard protocols, code obfuscation, dedicated hardware whose functionality is absent from dumped firmware, or steganography.

\paragraph{Type II: Compromised benign vehicle}
A \textit{Type II} vehicle originates as a benign product from a manufacturer without malicious intent.
Its integrity is compromised through a security breach, such as a malicious supply-chain component, software infection, vulnerability exploitation, compromised maintenance operation, or post-deployment intrusion \cite{AutomotiveSecuritySurvey1, AutoSWSuppChainNDSS, HCRLiviDetection}.
The adversary can exploit the vehicle's genuine status in an attack, but the extent of its control is determined by the affected components, available privileges, and reachable internal interfaces.
A sufficiently extensive compromise may approximate whole vehicle control; 
nevertheless, this configuration differs in origin from \textit{Type I} because the adversary must overcome a system originally designed without its cooperation.
Conventional automotive malware detection, intrusion detection, secure update, and network segmentation research primarily addresses this configuration.

\paragraph{Type III: Reappropriated genuine component}
An adversary can obtain a genuine vehicle component---for instance, from a scrapped vehicle, spare parts stock, or theft---and operate the vehicle part outside the system to which it was assigned.
For example, an on-board unit (OBU), communication module, or credential-bearing controller salvaged from a decommissioned vehicle may retain valid identifying material and be reappropriated to impersonate the donor vehicle or inherit an associated role, such as that of an incident response vehicle.
In another case, even where its identity has been revoked or is absent, the normal functionality of the component can be leveraged in adversarial activity.

For instance, Wouters et al. repurposed and modified a genuine Body Control Module (BCM) of Tesla vehicles as part of an attack against the Model X keyless entry system \cite{TeslaXSpareECUHack}.
The modified BCM was operated outside the affected vehicle to wake its paired key fob; the researchers then exploited the key fob to obtain a valid unlock token, opened the vehicle, and paired a modified key fob through the vehicle's diagnostic interface which is a procedure normally performed by a Tesla service technician.
Although the key fob and BCM employed certified Secure Elements and standard cryptographic primitives, implementation flaws in the firmware update and pairing protocols allowed these protections to be bypassed.
Component level authentication or physical fingerprinting would still recognise such hardware as genuine, even though its use is not authorised for the surrounding entity or operational context.
Detecting this configuration therefore requires verification of component-to-system binding, operational context, or behaviour in addition to component authenticity.

\paragraph{Type IV: Imitated vehicle or component}
An adversary may reproduce expected signals or protocol behaviour using non-genuine equipment including a software defined radio (SDR), field-programmable gate array (FPGA), commodity transmitter or purpose-built emulator.
Unlike \textit{Type III}, the adversary does not rely on an authentic component; it attempts to imitate its identity or observable behaviour.
Countermeasures for this type therefore extend beyond credential and protocol validation to physical characteristics associated with genuine hardware.
RF fingerprinting, for example, has been proposed to distinguish legitimate key fobs from imitators \cite{emsecRFfingerprinting}.
Masquerading, spoofing, message fabrication and Sybil attacks considered in vehicular network research commonly fall within this category \cite{VANETSecuritySurvey1, MisbehaviourDetectionSurvey}.

\textit{Types II--IV} differ in whether the adversary controls a genuine vehicle, a genuine component, or an imitation.
They nevertheless share an important assumption: some independent manufacturer, component, credential, or reference behaviour remains available as a basis for establishing legitimacy.
\textit{Type I} removes this assumption.
Authenticity verification may confirm that the vehicle contains manufacturer approved hardware and software, while behavioural monitoring may observe no violation at its cooperative interfaces.
Existing mechanisms may therefore produce no evidence until the hidden functionality is activated, and covert sensing or collection may never appear as vehicular network misbehaviour.

\begin{table*}[t]
	\caption{Behaviour--Locus Matrix with representative actions across operational loci and behavioural roles.}
	\label{tab:behaviour-matrix}
	\centering
	\footnotesize
	\setlength{\tabcolsep}{5pt}
	\renewcommand{\arraystretch}{1.20}
	\begin{tabularx}{\textwidth}{@{}p{0.15\textwidth}XXX@{}}
	
	\toprule
	\textbf{Operational loci} &
	\textbf{Covert action}	 &
	\textbf{Overt action}	 &
	\textbf{Concealment action} \\
	\midrule

	\textbf{Hardware locus} &
	Undeclared passive environmental sensing; activation of a concealed peripheral; acquisition through a camera, microphone, or other non-emitting sensor. & 	%covert
	Radio-frequency or optical emissions; active sensing; unsafe or unauthorised physical actuation. &												%overt
	Concealment within the vehicle interior or an unrelated component; tamper triggered evidence erasure or self-destruction; fragile physical construction that impedes disassembly or causes evidence damage. \\	%conceal
	\midrule

	\textbf{Software locus} &
	Dormant trigger evaluation; target or context recognition; undisclosed processing or storage of collected data. & 							%covert
	Safety-constraint bypass; execution of unauthorised control actions; destructive or disruptive modification of externally observable system state. &	%overt
	Code obfuscation; detection of investigative environments; falsification or deletion of logs; disabling, deleting, or overwriting hidden functionality.  \\	%conceal
	\midrule
	
	\textbf{Network locus} &
	Passive monitoring of nearby or internal traffic; reception of broadcast identifiers; listening for potential communication endpoints without transmitting. &	%covert
	Attempts to join or establish a network connection; transmission of messages or data; disruption of an internal or external network. &				%%overt
	Blending undisclosed traffic with legitimate telemetry; suppressing communication when investigation is detected; switching to a fallback channel; notifying an external controlling party. \\ %conceal
	\bottomrule
	\end{tabularx}
\end{table*}

\subsection{Behaviour--Locus Matrix}
\label{sec:blm}

Investigating suspected hidden or malicious-by-design functionality requires systematic coverage rather than ad-hoc inspection of an apparently relevant component or interface.
A malicious design authority or manufacturer may distribute such functionality across cooperating hardware, software, and network elements, allowing each element to appear benign when examined in isolation.
The \textit{Behaviour--Locus Matrix} therefore decomposes the investigative search space according to two dimensions: operational locus and behavioural role.
Drawing upon the behaviour-oriented perspective described in Section~\ref{sub:malwareBehaviour}, the matrix serves as a coverage scaffold for identifying where relevant actions may be implemented or manifested and what role they may perform.

Each dimension comprises three categories: \textit{operational locus} distinguishes hardware, software, and network loci, whereas \textit{behavioural role} distinguishes covert, overt, and concealment actions.

\subsubsection{Operational Locus}

The first dimension identifies the operational locus in which an action is implemented or observed.
Unlike conventional computational systems, a vehicle is a distributed CPS comprising embedded computing platforms, communication systems, sensors, actuators, and mechanical assemblies.
The model consequently distinguishes hardware, software, and network loci.

\paragraph{Hardware Locus}
The hardware locus encompasses both the computational and physical elements of the vehicle.
Its computational aspect includes electronic control units (ECUs), sensors, actuators, storage devices, communication modules, cables, diagnostic interfaces, and physical communication ports.
Its physical aspect includes mechanical and electromechanical assemblies such as the powertrain, braking, steering, and suspension systems.

A design authority with control over the product architecture can incorporate hardware resources or interfaces that are not required for the vehicle's documented operation.
For example, a telematics control unit may contain additional flash storage, an undocumented communication module, or an embedded subscriber identity module (eSIM) capable of holding an undeclared connectivity profile.
Such a feature is not malicious merely because it is unused or undocumented; however, it becomes relevant when it supports concealed storage, communication, activation, or future deployment of hidden functionality.
The hardware therefore need not be optimised solely for documented functions: additional resources may be incorporated during design and reserved for later activation.

\paragraph{Software Locus}
The software locus includes bootloaders, firmware, real-time and general-purpose operating systems, applications, services, learned models, configuration data, stored data, and runtime state.
At lower software layers, controllers such as body control modules and other ECUs commonly execute firmware or real-time operating systems that directly manage sensors, actuators, and in-vehicle communication.
At higher layers, capable nodes such as telematics control units (TCU) and in-vehicle infotainment (IVI) systems may run general-purpose operating systems, execute multiple services or applications, and employ machine-learning models for perception or decision making.

In a malicious-by-design vehicle, functionality may consequently be distributed across several software layers.
Applications or firmware may collect and retain data, monitor for an activation condition, recognise an investigative environment, enumerate nearby identifiers such as Wi-Fi service set identifiers (SSIDs), or transfer files and messages to a specified external endpoint.
A component that appears benign in isolation may therefore participate in hidden behaviour only when combined with particular data, messages, vehicle states, or actions performed by other components.

\paragraph{Network Locus}
The network locus encompasses communication within the vehicle and between the vehicle and external entities.
External communication may use cellular networks such as LTE or 5G, Wi-Fi, Bluetooth, VANET communication, USB-connected devices, or manufacturer-operated connected services.
Intra-vehicle communication may use Controller Area Network (CAN), Local Interconnect Network (LIN), FlexRay, Automotive Ethernet, or proprietary links between components.

A malicious-by-design vehicle may passively scan and record nearby network information, attempt to establish an undisclosed connection, transmit collected data, disrupt communication, or receive instructions that activate pre-positioned functionality.
Undisclosed traffic may be carried through a dedicated channel or piggybacked on legitimate telemetry and connected-service communications.
Encryption, obfuscation, or application-layer fragmentation may further reduce the ability of an external observer to distinguish such traffic from ordinary vehicle operation.
A manufacturer-controlled backend may also retain the ability to update, delete, reconfigure, or otherwise manage vehicle software despite publicly representing that no such remote capability is available.

Even when cellular connectivity is apparently absent, a vehicle could use another available interface.
For example, an embedded Wi-Fi module could scan for nearby access points, associate with an open network, test Internet reachability, and transmit retained data to a remote endpoint.
A manufacturer-provided application installed on a cellular-enabled mobile device paired with the vehicle could enable the device to serve as a communication relay, forwarding vehicle data over the mobile network.
Manufacturer-operated or authorised service centres could also use diagnostic tools to retrieve data cached in the vehicle without disclosing that collection to the owner.
Within the vehicle, a gateway or controller may similarly remain unresponsive until it receives a particular message, require a challenge--response exchange, or selectively forward commands according to the recognised sender or receiving ECU.
Although these mechanisms also have legitimate security uses, an adversarial design authority could repurpose them to conceal communication or restrict activation to authorised controlling entities.

The three loci are not mutually exclusive.
For example, a communication port is a physical hardware artefact, its controlling firmware belongs to the software locus, and the traffic carried through it is observed in the network locus.
A single capability may therefore produce actions and evidence across several loci.

\subsubsection{Behavioural Role}
The second dimension describes the role performed by an action.
A \textit{covert action} advances a potential adversarial objective without producing an apparent unauthorised effect at the legitimate interface visible to a specified observer.
An \textit{overt action} produces a manifest unauthorised effect or an externally observable deviation from expected operation.
A \textit{concealment action} reduces the discoverability, attribution, analysis, or reproducibility of another action, capability, or associated evidence.

Covert and overt actions directly advance a potential adversarial objective, whereas concealment actions hide, preserve, or protect the functionality through which that objective is pursued.
Whether an action is covert or overt is observer-dependent: an action hidden from nearby vehicles or ordinary diagnostic interfaces may nevertheless be observable through independent instrumentation.

\paragraph{Covert Behaviour}
Covert behaviour advances a potential adversarial objective without causing an immediately apparent failure or unauthorised effect at the interface available to a specified observer.
Examples include environmental or occupant sensing, collection and storage of internal vehicle data, recognition of targets or contextual conditions, passive reception of messages, and preparation of data for later transmission.
A covert action may constitute the objective itself or serve as an intermediate step towards a subsequent action.

A vehicle performing such actions may continue to provide its advertised services and remain conformant at its legitimate communication interfaces.
The privacy concerns outlined in \S~\ref{sec:introduction} illustrate how extensive sensing and data collection may remain operationally inconspicuous despite occurring without meaningful notice or consent.
Consequently, conventional misbehaviour detection may observe an apparently normal vehicle while failing to recognise that its sensing, computation, storage, or communication resources are being used for an undeclared activity.

\paragraph{Overt Behaviour}
Overt behaviour produces an observable effect that conflicts with expected vehicle operation.
In a CPS, this effect may occur in either the cyber or physical domain.
Examples include manipulation or destruction of data, interference with communication, disabling of protective functions, bypassing of safety constraints, execution of unauthorised actuator commands, and deliberate alteration of the vehicle's physical state or motion.
The resulting disruption or damage may constitute the objective itself, as in sabotage, or may support a further objective.
Although overt behaviour is more readily observable than covert behaviour, a one-time or safety-critical physical effect may cause harm before conventional monitoring can detect and respond to it.

\paragraph{Concealment Behaviour}
Concealment behaviour obstructs the discovery, observation, analysis, attribution, or preservation of another capability or action.
Relevant behaviours include code obfuscation, anti-debugging and anti-analysis mechanisms, recognition of instrumented or investigative environments, delayed or context-dependent activation, falsification or deletion of logs, suppression of communication, and alteration of behaviour during inspection.

Concealment may also destroy or replace executable evidence.
For example, functionality may overwrite its malicious code with a benign substitute when investigation is detected, leaving the vehicle operational while removing the code that would otherwise support subsequent analysis.
Other actions may damage evidence-bearing hardware, notify an external controlling entity, or switch to a fallback communication channel.
These actions need not directly advance the primary objective; their role is to preserve the hidden functionality or reduce the likelihood that it can be discovered and reproduced.

Combining operational locus and behavioural role produces the nine cells shown in Table~\ref{tab:behaviour-matrix}.
The \textit{Behaviour--Locus Matrix} should be understood as an investigative coverage matrix rather than a taxonomy of nine mutually exclusive threat types.
It classifies actions rather than components or capabilities, and superficially similar behaviour may receive different classifications depending on its implementation and the observations available to a specified observer.

Consider data exfiltration through cellular communication.
Data collected without authorisation may be transmitted by piggybacking on legitimate telematics traffic.
The transmission constitutes an objective-serving network action, while piggybacking it on ordinary traffic performs a concealment role.
If the vehicle instead contains an undeclared and physically hidden cellular module, concealing that module constitutes a hardware concealment action, whereas transmitting through it constitutes a network action.
Whether the transmission is covert or overt depends on the observer's awareness of the vehicle's expected connectivity and capability to detect and attribute its radio-frequency emissions.
Therefore, a single action may occupy more than one cell when its implementation or observable effects span several operational loci.
Likewise, a single capability may be realised through coordinated actions that perform different behavioural roles.

\subsection{Behavioural Interaction Model}
\label{sec:bim}

\begin{figure}[t]
	\centering
	\includegraphics[width=\linewidth]{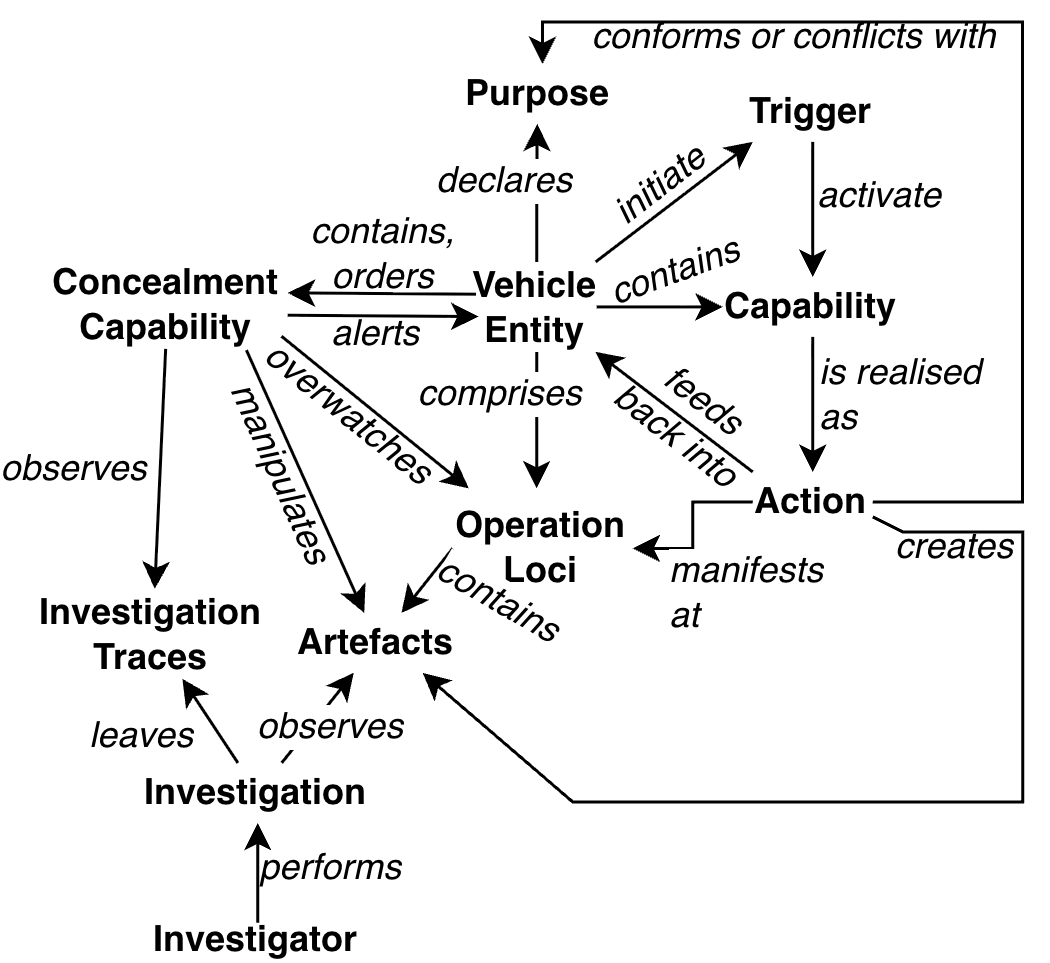}
	\caption{Conceptual Behavioural Interaction Model showing capability activation, action manifestation across operational loci of the Behaviour-Locus Matrix, artefact generation, investigation, concealment, and comparison with declared purpose.}
	\label{fig:bim}
\end{figure}

The \textit{Malign Entity Archetypes} describe the adversarial configurations considered by the threat model, while the \textit{Behaviour--Locus Matrix} (BLM) classifies actions according to their behavioural roles and operational loci.
The \textit{Behavioural Interaction Model} (BIM) complements these components by representing how vehicle capabilities are activated and realised as actions, how those actions manifest and leave artefacts, and how concealment may interfere with their investigation.
Figure~\ref{fig:bim} presents these relationships as a conceptual model rather than a chronological attack sequence or an exhaustive catalogue of behaviours.

\subsubsection{Capability Activation and Action}
A vehicle entity contains capabilities realised through its hardware, software, and network elements.
A capability describes what the entity is able to do, regardless of whether that functionality is declared, hidden, continuously exercised, or currently dormant.
The capability itself is not assigned a behavioural role; that classification applies to the actions through which it is exercised.

A trigger represents the activation condition or operating mode under which a capability is exercised.
It may consist of a timer, an inbound command, a recognised location or target, a particular vehicle state, a combination of contextual conditions, or continuous operation without a discrete activating event.
The trigger may therefore be internally initiated or evaluated in response to an external condition.
Once activated, the capability is realised through one or more actions.

The same capability may support actions having different behavioural roles.
For example, a network capability may be used covertly to monitor nearby traffic or overtly to disrupt a network.
Likewise, control over an active sensor may support its declared sensing function, undeclared collection, or an overt action that repurposes its emissions to interfere with another system.
The classification consequently applies to the realised action rather than to the underlying capability alone.

An action is classified as \textit{covert} when it advances a potential objective without producing an immediately apparent unauthorised effect at the interface available to a specified observer.
Examples include passive monitoring, undeclared sensing, contextual recognition, data retention, and transmission piggybacked on legitimate communication.
An action is classified as \textit{overt} when its occurrence or effects are perceptible to a specified observer, either directly through human senses or through available sensing, monitoring, or measurement equipment.
Whether an action appears covert or overt can depend on the observer's awareness and technical capability: behaviour hidden from peers or ordinary monitoring may be apparent to an investigator using specialised instrumentation.

Every action also manifests at one or more operational loci within the vehicle.
Crossing its behavioural role with its hardware, software, or network locus places the action within the corresponding BLM cells.
The operational loci thus connect the abstract capability--action relationship to the concrete components and interfaces through which the behaviour occurs.
An action may additionally feed back into the vehicle by altering its internal state, supplying further decision inputs, or enabling subsequent actions.

\subsubsection{Concealment and Investigation}
Actions may create artefacts within the operational loci at which they manifest.
These artefacts may include stored logs, network traffic, electromagnetic emissions, and physical or electrical changes to a component or its environment.
An investigator performs an investigation by attempting to acquire and interpret such artefacts.
Not every artefact is necessarily retained, accessible, or recognised, and a single action may leave different artefacts across several loci.

The investigation may itself leave traces within the vehicle.
Diagnostic requests, test commands, connected instruments, absence of peer components, and physical disassembly may reveal that the vehicle is being examined.
A concealment capability may observe these traces and alert the vehicle, allowing other capabilities to suppress, postpone, or modify their actions while investigative conditions persist.

Concealment capabilities may also monitor relevant operational loci and manipulate the artefacts available to the investigator.
Such manipulation may include deleting or falsifying logs, altering stored data, suppressing communication, overwriting hidden functionality with a benign substitute, or presenting different behaviour under test conditions.
Although shown separately in Figure~\ref{fig:bim} to foreground its investigative significance, a concealment capability remains a capability realised through actions.
Those actions can themselves be classified within the BLM and may leave further artefacts.

Investigation and concealment therefore form an interactive relationship.
The investigator changes the conditions surrounding the vehicle, while the vehicle may detect and respond to those changes.
Consequently, failure to observe a suspected action does not establish the absence of the corresponding capability, particularly where activation is contextual or investigation-aware concealment is possible.

\subsubsection{Purpose Assessment from Observed Behaviour}
The vehicle entity declares a purpose against which its realised actions may be assessed.
This declaration provides a reference for investigation but is not presumed to constitute a complete or necessarily truthful account of the vehicle's implemented functionality.
An individual action may conform to the declared purpose, extend beyond it, or conflict with it.
The assessment must therefore consider the relationships among capabilities, triggers, actions, operational loci, and the artefacts supporting their reconstruction.

Accumulated observations may support an inference about the vehicle's \textit{evidenced operational purpose}: the objective that its implemented capabilities and actions appear structured to serve.
This does not claim direct knowledge of the design authority's subjective intention.
Rather, it enables comparison between declared purpose and behaviour supported by available evidence.
Neither a single unusual action nor an unobserved suspected action is sufficient on its own to establish this relationship.

The \textit{Behavioural Interaction Model} thereby materialises the \textit{Behaviour-Locus Matrix} by relating its classified actions to the capabilities, activation conditions, artefacts, and concealment mechanisms from which they arise.
It also connects the behavioural analysis back to the first \textit{Malign Entity Archetype}: a vehicle may remain authentic, protocol-conformant, and apparently benign to its peers while its evidenced operational purpose conflicts with the purpose declared for it.

\section{Discussion}
Peer-benign, purpose-malicious functionality can undermine justified trust in the service delivered by a CPS, affecting availability, reliability, safety, integrity, and confidentiality~\cite{AvizienisEtAl2004}.
This breadth underscores the need for assurance that does not presume a benign design authority.

\subsection{Security Controls under an Adversarial Design Authority}
Conventional vehicular cybersecurity seeks to protect vehicles against unauthorised access, modification, and control.
Accordingly, security mechanisms restrict access to vehicle software, diagnostic interfaces, cryptographic material, update functions, and internal communication, while research examines how attempted intrusion can be prevented, detected, and mitigated.

Under the malicious-by-design assumption, however, the design authority controls many of these protective mechanisms.
The same controls that defend a vehicle against external attackers may restrict independent access, conceal undocumented interfaces, suppress relevant observations, or permit examination only through manufacturer-controlled tools.
Independent investigation may consequently require investigators to bypass protections in ways that are technically similar to an intrusion.
The problem is not that these protections are inherently undesirable, but that their trust boundary includes the party whose functionality is under examination.

The \textit{Type III} in the \textit{Malign Entity Archetypes}, the reappropriated genuine component, presents a related tension between security and repairability.
Component-to-vehicle binding can restrict the unauthorised reuse of genuine components that retain valid identities or useful capabilities.
However, strict pairing may also disable genuine replacement parts and restrict independent repair or owner maintenance.
Simple environmental checks may be circumvented, whereas stronger binding may reduce misuse at the cost of increased dependence on manufacturer-controlled tools and services.
Such mechanisms must therefore distinguish unauthorised reappropriation from legitimate maintenance rather than treating operation outside the original configuration as inherently malicious.

\subsection{Limitations}
This work presents a conceptual threat model rather than an operational investigation or detection method. 
The \textit{Malign Entity Archetypes}, \textit{Behaviour--Locus Matrix}, and \textit{Behavioural Interaction Model} structure the relevant threat configurations, investigative coverage, and behavioural relationships, but do not prescribe evidence acquisition procedures, analytical tools, or decision thresholds.
The proposed model has also not yet been empirically evaluated using vehicles or testbeds containing seeded hidden functionality.

The behaviours and manifestations presented in the \textit{Behaviour--Locus Matrix} are illustrative rather than exhaustive, and a single capability may produce actions and evidence across several matrix cells.
Classification may also depend on the specified observer: behaviour concealed from peers or ordinary diagnostic interfaces may remain observable through independent instrumentation.
Nevertheless, dormant, context-triggered, or concealment-protected functionality may evade available observations.
Consequently, the absence of observed behaviour cannot establish the absence of the corresponding capability.
Finally, the model can characterise functionality and evidence consistent with an adversarial objective, but cannot by itself establish subjective manufacturer intent, determine the prevalence of such functionality, or assign legal responsibility.

\subsection{Future Work}
Future work may further formalise the elements and relationships of the \textit{Behavioural Interaction Model} and clarify how represented actions relate to its cells. 
A subsequent extension could develop a tool-supported investigative workflow combining static examination, dynamic observation, and contextual stimulation while considering evidence sources independent of the suspected design authority. 

Another direction is to examine how concepts from the \textit{Malware Behavior Catalog} and \textit{Malware Attribute Enumeration and Characterization} could be systematically adapted to cyber-physical systems.
Such an adaptation could incorporate physical actions, contextual triggers, distributed components, environmental interactions, and cyber-physical observables, providing a basis for further elaboration of the proposed model.
Other possible directions include independent logging, randomised assessment, additional assurance mechanisms, and application of the model to CPS domains beyond vehicles.

\subsection{Conclusion}
This paper examines vehicle entities whose components and communications may remain authentic and protocol-conformant while their implemented functionality violates their declared or authorised purpose.
To characterise this threat, we formulate an \textit{Entity-Centric Threat Model} comprising three complementary components.
The \textit{Malign Entity Archetypes} distinguish malicious-by-design, compromised, reappropriated, and imitated entities; 
the \textit{Behaviour--Locus Matrix} classifies overt, covert, and concealment actions across hardware, software, and network loci; 
and the \textit{Behavioural Interaction Model} relates capabilities, triggers, realised actions, artefacts, investigation, concealment, and declared purpose.

Together, these components provide a conceptual foundation for investigating functionality that may be distributed, contextually activated, and concealed from peers or conventional monitoring.
They demonstrate that authenticity, component genuineness, and protocol conformance do not alone establish the trustworthiness of a vehicle's operational purpose.
Automotive assurance must therefore account for the possibility that the design authority itself lies within the threat boundary.

\bibliographystyle{IEEEtran}
\bibliography{references}
\end{document}